\documentclass[letterpaper, 10 pt, conference]{ieeeconf}  

\IEEEoverridecommandlockouts                              

\usepackage{cite}

\usepackage[pdftex]{graphicx}
\graphicspath{{Figures/}{Figures/ExampleImages/}{Figures/CNNs/}}
\DeclareGraphicsExtensions{.pdf,.jpeg}
\usepackage{amsmath, amsfonts}
\usepackage{algorithmic}

\usepackage{caption}
\usepackage{subcaption}

\usepackage{url}

\newcommand{\includeCroppedPdf}[2][]{%
    \IfFileExists{./#2-crop.pdf}{}{%
        \immediate\write18{pdfcrop #2 #2-crop.pdf}}%
    \includegraphics[#1]{#2-crop.pdf}}

\renewcommand{\vec}[1]{\mathbf{#1}}

\title{\LARGE \bf
Enhanced Transfer Learning Through Medical Imaging and Patient Demographic Data Fusion
}

\author{Spencer A. Thomas$^{1}$ 
    \thanks{$^{1}$Spencer A. Thomas is with the Data Science group, National Physical Laboratory, Teddington, UK
        {\tt\small spencer.thomas@npl.co.uk}
    }%
}

\begin{document}

\maketitle
\thispagestyle{empty}
\pagestyle{empty}


\begin{abstract}
In this work we examine the performance enhancement in classification of medical imaging data when image features are combined with associated non-image data. We compare the performance of eight state-of-the-art deep neural networks in classification tasks when using only image features, compared to when these are combined with patient metadata. We utilise transfer learning with networks pretrained on ImageNet used directly as feature extractors and fine tuned on the target domain. Our experiments show that performance can be significantly enhanced with the inclusion of metadata and use interpretability methods to identify which features lead to these enhancements. Furthermore, our results indicate that the performance enhancement for natural medical imaging (e.g. optical images) benefit most from direct use of pre-trained models, whereas non natural images (e.g. representations of non imaging data) benefit most from fine tuning pre-trained networks. These enhancements come at a negligible additional cost in computation time, and therefore is a practical method for other applications.

\end{abstract}




\section{Introduction}

In medical applications such as diagnosis from medical images, clinicians will typically have access to relevant information that is not contained within the image. A patient's clinical or demographic history may have a significant role in the clinical decision making pipeline. For instance, the age of a patient may significantly affect the probability or risk of a disease and hence may increase the weighting on specific features or borderline instances. Other cases may not be as obvious or explicitly known to the clinician. In any case, analysis based on the imaging data alone does not fully utilise relevant data available and limits the potential of AI-assisted decision making in biomedical applications.

Deep learning has emerged as a powerful suite of tools for image classification \cite{LeCun2015}, and has a huge potential to solve challenges in healthcare settings. The use of deep neural networks is successful at tasks such as classification of medical images \cite{litjens_survey_2017}, analysis of electronic health records \cite{Thomas2019CMR, deLusignan2019, Avati2018} and segmenting data from emerging medical technologies \cite{Thomas2017, Behrmann2018}. This enormous potential comes with the caveat that very large amounts of data are required to train robust models that generalise beyond the training set. This requirement is unfortunately difficult to satisfy in the majority of biological and medical studies due to barriers to data availability.


Transfer learning has emerged as a promising method for circumventing the need for vast amounts of data to train deep networks \cite{Lakhani2018}. For domains with limited data, transfer learning utilises networks pre-trained on similar tasks with large amounts of data \cite{Pan2010}. Transfer learning is often used in medical imaging \cite{Xu2017, Litjens2017, Shin2016} due to the limited availability of data that require expert labeling \cite{Rai2019transfer}. Transferring the image features from one domain to another can at least match the performance of models trained directly on the new domain \cite{tajbakhsh_convolutional_2016}, though it is not known if this is a general property as some cases the performance is worse than hand crafted features \cite{Aston2019}, 
Moreover, the configuration of the transfer can be performed in a number of ways \cite{Rai2019transfer, Rai2019aggreated} and more research is needed in this area.

Medical imaging data often has associated metadata used by clinicians in patient assessments. These metadata are multi type (numeric, categorical, etc) and are essential for maintaining the value of archived data \cite{Smith2019}. The information may be content related, e.g. scanner parameters, or relevant extracts from computerised medical records (CMR). These resources contain rich information relating to diseases \cite{de_lusignan_rcgp_2017,correa_royal_2016}, and data driven methods can identify patterns of patients \cite{Thomas2019CMR,deLusignan2019}. 

Classification tasks based on the combination of imaging with genomics data has been shown to surpass clinical experts in digital pathology \cite{Mobadersany2018}. 
Combining relevant information about the sample, e.g. patient demographics, with imaging data can lead to higher accuracy in binary classification tasks \cite{rocheteau_deep_2020}. An enhancement in classification performance has been observed in transfer learning for specific configurations with a single dataset \cite{Thomas2021EMBS}, though it is unknown if this applies to other methods of transfer learning or target domains. Furthermore, the explain-able origin of an enhancement in performing from such a framework is needed.

Clinicians will typically base diagnosis on several information sources either implicitly or explicitly. Demographic factors such as age can influence the likelihood of disease prevalence. In this work we investigate the combination of imaging data with related metadata to enhance classification performance evaluated by several metrics. We utilise transfer learning due to the limited volumes of data available, comparing the performance with and without metadata. Additionally we repeat the experiments with and without data augmentation during the training of the model.

Metadata is needed for data curation and essential for maintaining the value of archived data \cite{Smith2019}. Metadata for medical applications contain vital information regarding provenance and parameters of acquisition that could impact data processing or interpretation of results. 
The potential of metadata for healthcare data not only lies in the curation and knowledge of the data, but it can be a rich source of information itself. Computerised medical records contain a patient (meta) data and are often complex high dimensional data sources. Methods for analysing CMRs have uncovered low dimensional patterns relating to sub-groups of patients \cite{Thomas2019CMR,deLusignan2019} and have enormous potential for insight into biomedical sciences. This paper focuses on the potential of integrating imaging data with non-imaging data (e.g. metadata, CMRs, etc) as a means to improve performance of classification of medical imaging. The value for information within non-imaging data may enhance the predictive power of deep learning classification methods through the combination of relevant information with traditional image features extracted by a deep network. These multi-modal data can be combined prior to the classification in a deep network offering a new direction of research in how to combine these data to most improve model performance. 

Here we investigate how the inclusion of non-imaging data affects the performance of transfer learning based multi-class classification problems using real world data. For generality, we consider the ISIC \cite{ISIC} skin imaging and image representations of the PTB XL \cite{PTBXL} ECG datasets. We perform experiments with eight popular convolutional neural network models to investigate the effects of combining imaging and non-imaging data on the classification performance assessed via several metrics. We conduct experiments for each network with and without image augmentation. Finally, we repeat these experiments using network weights obtained directly from training on the ImageNet \cite{ImageNet} dataset, retraining the classification layer only (so called {\em bottleneck feature extraction}), and compare to retaining the ImageNet weights using the target dataset (so called {\em fine tuning}).

\section{Methods}
For imaging data $X \in \mathbb{R}^{N \times D}$ where $X_i$ is a $D$ dimensional data point with $N$ images, deep learning applies a series of transformations through hidden layers typically in the form of convolutions. These transformations occur through a series of layers in the network which ultimately reduce the dimensionality of the input data into a feature vector of dimensionality $d_k$ where $d_k < D$.  
At each layer, following the notation of \cite{Vidal2017}, a matrix $W^k \in \mathbb{R}^{d_{k-1} \times d_k}$ linearly transforms the output of the $(k-1)$th layer, $X_{k-1} \in \mathbb{R}^{N \times d_{k-1}}$, into a $d_k$ -dimensional space, $X_{k-1}W^k\in \mathbb{R}^{N \times d_{k}}$, at the $k$th layer. The linear transformations are followed by a non-linear activation function, $\sigma_k(z)$, here a rectified linear unit (ReLU), at each layer. The output of a network with $K$ layers is given by
\begin{equation}
\label{eq:deepLearning}
\mathcal{F} \left( X \right) =
\sigma_K \left( \dots 
\sigma_2 \left( \sigma_1 \left( X W^1 \right) W^2 \right) \dots W^K \right)~,
\end{equation}
where $W^k$ represents the weight matrix at layer $k$ and the bias vectors are omitted for brevity. 

Clinicians will typically have access to information that is not contained within images that is potentially relevant to biomedical tasks. We refer to this additional non-imaging data as metadata throughout this paper to reflect the generality of this information, which can relate to the patient and/or acquisition of the data. In order to utilise metadata associated with images, we use the framework outlined in \cite{Thomas2021EMBS} to combine the image features obtained from a deep neural network and the associated metadata. We use a method inspired by \cite{Thomas2019CMR} to obtain representations of the metadata that are compatible with the image features obtained in the neural network. This maps the metadata $M$ to a numerical vector $\mathcal{G} \left(M \right)$ 
\begin{equation}
\label{eq:G}
\mathcal{G}\colon\mathbb{M}\!\!\to\mathbb{M^+}
\end{equation}
where $\mathbb{M^+}$ is $\mathbb{M}$ mapped to ASCII decimal described as follows. All numerical data are parsed directly. Chronological information is parsed with a format of `yyyy-MM-dd hh:mm:ss', then separated into a 1x6 numeric array, e.g. [2001; 05; 28; 12; 49; 25]. All other non numeric data are grouped into categories and then replaced with a category index. Any not a number entries in the data are replaced with the minimum value -1 to differentiate it from the rest of the data. The metadata is integrated with the image data by concatenating the image features, $\mathcal{F}(X)$, with the mapped metadata $\mathcal{G}(M)$, formally, 
\begin{equation}
\label{eq:featureFusion}
\begin{split}
\mathcal{H} &= 
\begin{pmatrix} 
\mathcal{F}(X) \quad \mathcal{G}(M) \\
\end{pmatrix} 
\\
&= 
\begin{pmatrix} 
\mathcal{F}_{1,1} & \cdots & \mathcal{F}_{1,d_K} & \mathcal{G}_{1,1} & \cdots & \mathcal{G}_{1,d_{K'}}\\ 
\vdots & \ddots & \vdots & \vdots & \ddots & \vdots \\ 
\mathcal{F}_{N,1} & \cdots & \mathcal{F}_{N,d_K} & \mathcal{G}_{N,1} & \cdots & \mathcal{G}_{N,d_{K'}}
\end{pmatrix} 
\end{split}~,
\end{equation}
where $N$ is the number of images, $d_K$ is the dimensionality of the output of the neural network, $\mathcal{F}(X)$, and $d_{K'}$ is the dimensionality of the converted metadata, $\mathcal{G}(M)$.


In this work we are interested in biomedical classification tasks using the features obtained from the data. As we are investigating the potential benefits of combining metadata with image features obtained from deep neural networks, we use a softmax classifier in our experiments. This ensures direct comparability between performance with and without the inclusion of metadata. Moreover, this allows comparison to standard neural networks which typically employ a softmax classifier for image classification tasks. The softmax function, a generalisation of logistic regression, builds a classification model for $K$ classes by optimising a weight vector $\vec{w}$ to maximise the the predicted probability $P$ that the input $\vec{z}$ is assigned to the correct class, 
\begin{equation}
\label{eq:softmax}
P\left( y=j | \vec{z} \right) = \frac{e^{\vec{z}^T \vec{w}_j}}{\sum_{k=1}^K e^{\vec{z}^T \vec{w}_k}}~.
\end{equation}
This classification model is trained using gradient descent for a maximum of 2000 epochs or when the gradient falls below 10$^{-6}$. The input to the softmax classifier, $\vec{z}$ is $\mathcal{F}(X)$ from Eq.~(\ref{eq:deepLearning}), or $\mathcal{H}$ from Eq.~(\ref{eq:featureFusion}).

\subsection{Experiments} 

In order to obtain the image feature vector $\mathcal{F}(X)$ from Eq.~(\ref{eq:deepLearning}) we utilise several state-of-the-art deep convolutional neural network architectures. As we are investigating transfer learning, all of the networks used have been pretrained using the ImageNet dataset \cite{ImageNet}. In this work we compare the following network architectures {alexnet} \cite{alexnet}, {mobilenetV2} \cite{mobilenetv2_2018}, {densenet201} \cite{Huang2016}, {resnet50} \cite{He2015}, {inceptionresnetv2} \cite{Szegedy2017}, {vgg16} and {vgg19} \cite{Simonyan2015} and googlenet \cite{googlenet}. 
The specific layer we extract the image features from varies with the different models, but in all cases we extract the feature vector at the deepest layer in the network (prior to classification) which typically corresponds to the lowest dimensionality representation of the input image data.

We conduct several experimental setups with these models, as follows. First we use the networks as feature extractors using the ImageNet optimised network weights, known as bottleneck feature extraction (BN) \cite{Rai2019transfer}. We compare the classification performance based on these BN features to the performance when combining these features with the associated metadata as in Eq.~(\ref{eq:featureFusion}). Secondly, we perform the same set of experiments fine tuning (FT) the ImageNet optimised weights with target data. In this configuration the network weights are retrained using images from the target domain by replacing the classification layer and training in a supervised manner. Once optimised, we compare the performance with and without the metadata as in the BN experiments above. Thirdly, we evaluate the performance in the above two experiments when using image augmentation during training or feature extraction in the BN experiments outlined as follows. The augmented image $X'$ is obtained from the function $\Omega$ applied to the input image, $x$, 
$X' = \Omega \left( X \right) $.
The augmentation function introduces a random shift in the image of up to 30 pixels from its origin separately along the x and y axes, random reflections in x and/or y, and random rotations up to 90 degrees. These transformations are applied to the training and testing data when using augmentations.

\subsection{Datasets}
\label{sec:data}
We apply our methods to two distinct large open source datasets to demonstrate the generality of the framework and improvements in different application domains. For simplicity of comparison we restrict these experiments to imaging domains, but note that the framework is readily applicable to non-imaging domains. We utilise the HAM1000 dataset from the International Skin Imaging Consortium (ISIC) database that contains 10,015 digital images of skin lesions with accompanying metadata for each image \cite{ISIC}. These metadata include patient demographics which may be relevant for disease classification. We also utilise the PTB XL electrocardio graph (ECG) dataset \cite{PTBXL} which contains 12-lead ECG recordings from 21,837 patients with accompanying metadata. Note that we convert the ECG signals into images to utilise pre-trained networks allowing direct comparisons with the ISIC data. Further details of both datasets are given in the following subsections.

\subsubsection{ISIC Dataset}
The 10,015 skin lesion images in the HAM1000 dataset belong to one of eight classes categorised by clinical experts \cite{ISIC}. The dataset is imbalanced; the number of instances in each class range from 6705 in the majority class to 115 in the minority class. 
Each of the 10,015 images in the HAM10000 dataset has an associated metadata file in {\it JSON} format with clinical information about the patient. These include demographic information (age and gender) and anatomical location of the lesion that are potentially diagnostically relevant. The patient metadata also include references to the diagnosis. With the exception of the diagnostic label for supervised training, this latter information is discarded to prevent bias in the model. 


\subsubsection{PTB XL Dataset}
The PTB XL data set contains 12 lead ECG signals from 21,837 patients that have been labelled by clinical experts \cite{PTBXL}. The labels for the patients have been grouped into five super-classes based on the pathology of the diagnosis; normal (NORM), Myocardial Infarction (MI), Conductive Disturbance (CD), Hypertrophy (HYP) and ST/T-Change (STTC) see \cite{PTBXL} for further details. 
Here we refer to the samples by their super-class labels from now on. 
We only consider the 15,351 samples with unique super-class labels in this study.



\begin{figure*}[t]
\centering
\includegraphics[width=0.95\textwidth]{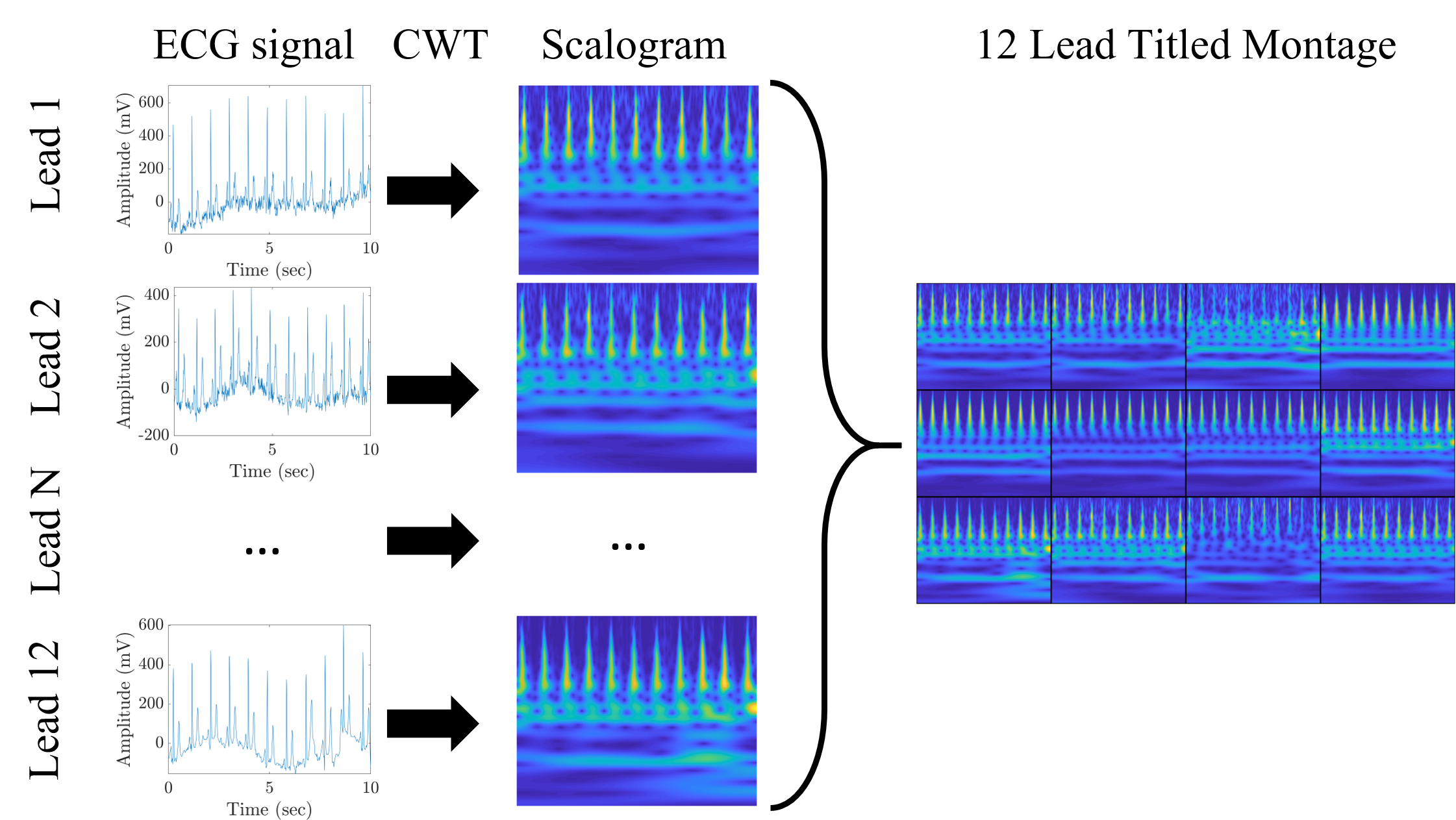}
\caption{Illustration of the 12 lead montage of scalograms generated from the PTB XL data prior to using the deep neural networks.}
\label{fig:leadMontage}
\end{figure*}

In order to utilise transfer learning through pretrained CNN networks, and compare results to the ISIC data, we represent the ECG data as time-frequency `images' called scalograms. These images have been shown to yield the best performance for transfer learning tasks on ECG data compared to other imaging representations of ECG data \cite{venton_signal_2020}. Scalogram images are generated by plotting the absolute value of the coefficients from a continuous wavelet transform (CWT) of the signal. The CWT of an ECG signal $x(t)$ at time $t$ is given by
\begin{equation}
    X_w(a,b) = \frac {1}{|a|^{1/2}} \int_{-\infty}^{\infty} x(t) \bar{\psi} \left( \frac{t-b}{a} \right) \delta t
\end{equation}
where $a \in \mathbb{R}^{+}$ is a positive scale parameter, and $b\in \mathbb{R}$ is a a translation parameter. The power spectrum, $|X_w(a,b)|^2$, can be represented for varying time and frequency images. 
We use a Morse wavelet with a symmetry parameter set to three and a Time bandwidth of 60. We select a sampling frequency of 100 to reflect the intervals in the data, with 10 `Voices Per Octave' for each lead in the ECG data.
In order to obtain a single image for each sample from the 12-lead ECG data, we combine the 12 generated scalograms for each sample into a single tiled 12 lead montage as illustrated in Fig.~\ref{fig:leadMontage}. 
This ensures all available information is being used by the deep learning model. Note that an assessment of utility of a given ECG lead or scalogram for classification is outside the scope of this work. 

The ECG data in the PTB XL dataset contain metadata for each patient, this is consistent across all 12 leads for a given patient. These include clinical information about the patient and diagnosis and information about the acquisition of the data. We do not consider the latter to avoid detection of potential bias in the data. The diagnostic information is used to determine the super-class label and the patient information is fused with the image data in our method. 



\subsection{Evaluation Metrics}
We evaluate the performance of our experiments via several metrics to provide a holistic view of these methods. 
We look at the Accuracy, Specificity, Sensitivity, Precision, F-measure, Informedness, Markedness and Matthews Correlation Coefficient (MCC), as well as the area under the receiver operator characteristic curve (AUROC). We consider the macro average for these metrics to provide a view on the multi-class problems. In order to observe the effect of fusing the metadata with the image features in the classifiers, we compute the percentage improvement (or degradation) of these metrics when incorporating the image features, compared to only using the image features. In addition to the macro average we also include the standard deviation across the classes for each metric. 
\section{Results}

\begin{figure*}[t!]
\centering
\includegraphics[width=0.99\textwidth]{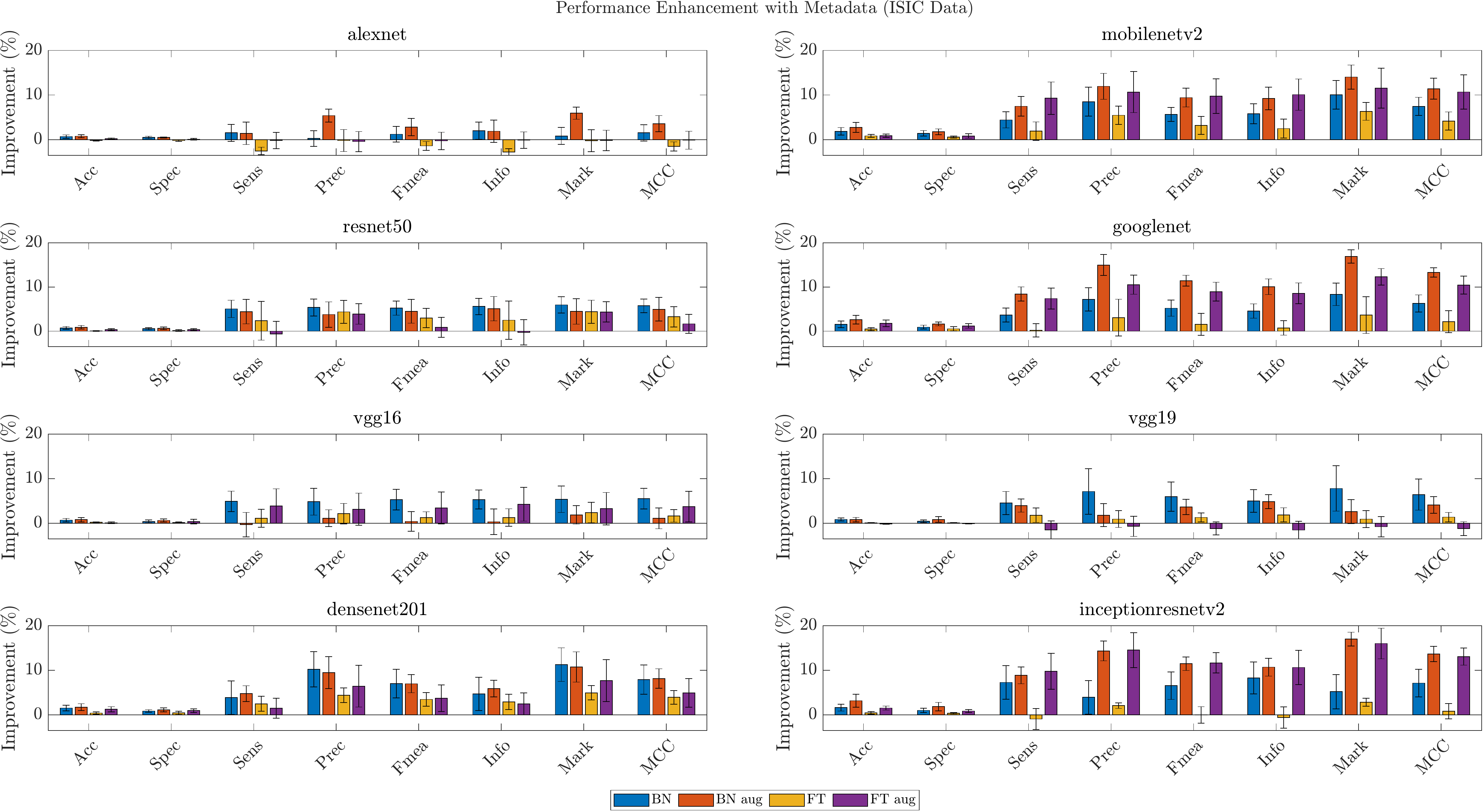}
\caption{Improvement in macro average performance of transfer learning in deep neural networks when using image metadata. Values are the difference in performance scores with positive values demonstrating improved performance when using metadata with image features. For example scores of 70$\%$ (image only) and 80$\%$ (combined image and metadata) would be plotted as 10$\%$. }
\label{fig:MacroAverageSKIN}
\end{figure*}

\subsection{Experimental Set up}
To account for the difference in input size to each network, all images are resized to the required dimensions using bi-linear interpolation. For the fine tuning experiments the networks are optimised using stochastic gradient descent with momentum, where momentum is set to 0.9 with an initial learning rate of $3\times10^{-4}$. The order of the training images is shuffled at each epoch. All the ISIC data are trained with a mini batch size of 128 on a 2.2GHz intel(R) Xeon(R) E5-2698 v4 CPU. We split the 10,015 skin images into 7,021 training, 1,502 validation, and 1,502 testing samples, maintaining the class distributions in each subset. All PTB XL data are trained with a mini batch size of 64 on an NVIDIA RTX 3090 GPU. We split these samples into 10,746 training, 2,303 validation, and 2,302 testing samples, maintaining the class distributions in each subset. All experiments were performed in MATLAB R2021a (V 9.10.0.1684407 Update 3).

We evaluate the models when using the pretrained models directly and fine tuning these pretrained networks. In both cases the final network is used to extract image features which are then; 1) used to train a softmax classifier, Eq.~(\ref{eq:softmax}); and 2) combined with the metadata features, Eq.~(\ref{eq:featureFusion}), before training a separate softmax classifier, Eq.~(\ref{eq:softmax}). In all cases the softmax classifier is trained with the training set (including the bottleneck configuration) and evaluated with the unseen testing data. The softmax classifiers are trained for a maximum of 2000 epochs or a minimum gradient of 10$^{-6}$. We repeat this set up when including image augmentation for direct comparability.




\begin{figure*}[t!]
\centering
\includegraphics[width=0.99\textwidth]{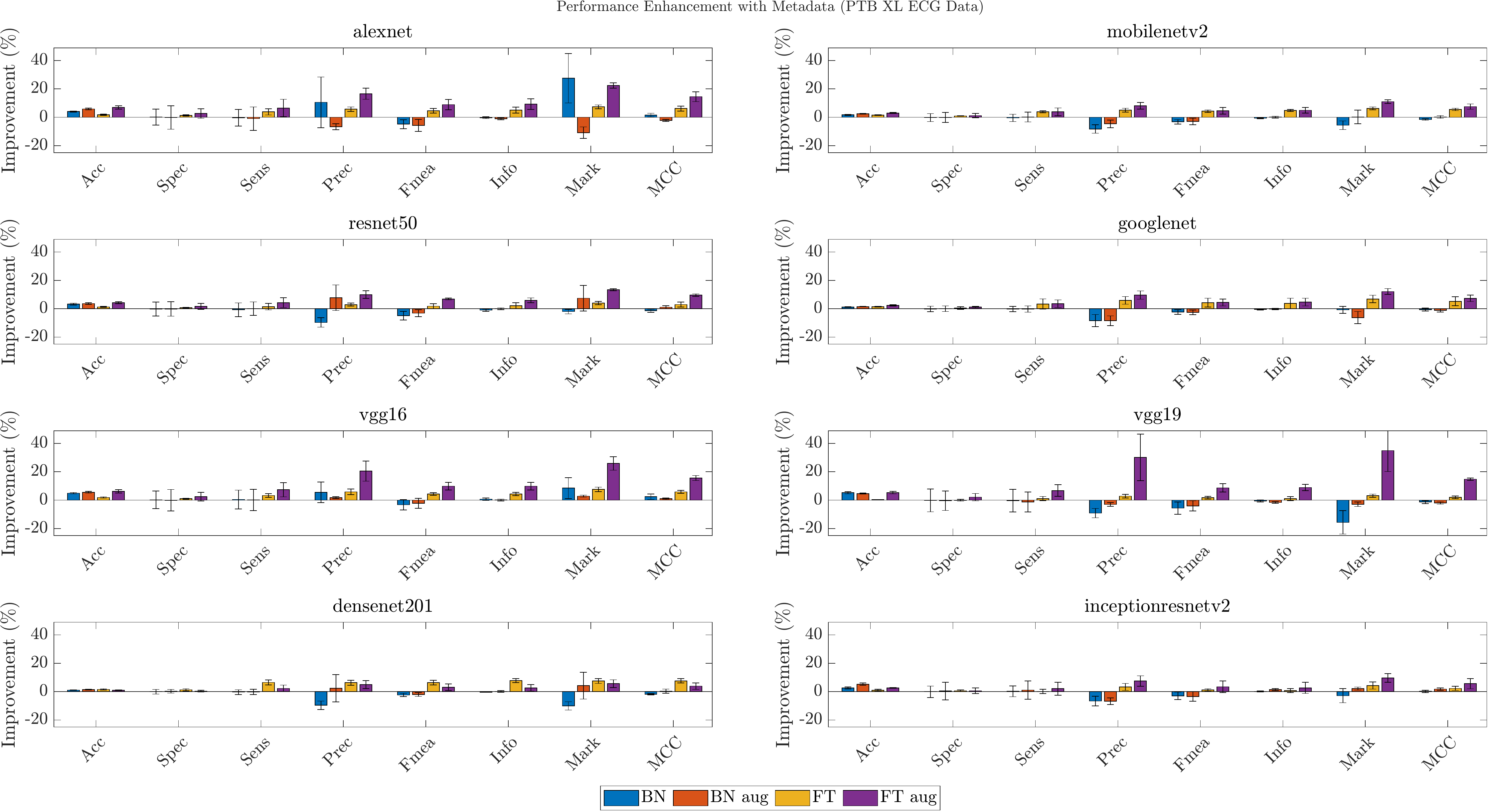}
\caption{Improvement in macro average performance of transfer learning in deep neural networks when using image metadata. Values are the difference in performance scores with positive values demonstrating improved performance when using metadata with image features. For example scores of 70$\%$ (image only) and 80$\%$ (combined image and metadata) would be plotted as 10$\%$. }
\label{fig:MacroAveragePTB}
\end{figure*}

\subsection{ISIC Performance}
All the networks tested demonstrated an improvement in model performance when incorporating metadata for almost all metrics as seen in Fig.~\ref{fig:MacroAverageSKIN}. MobileNetV2, GoogleNet, Densenet201 and InceptionResNetV2 demonstrated particularly large improvements when incorporating the metadata. Interestingly, the performance increase is typically lower in these models when fine tuning without augmented images in these networks. This maybe a result of the relatively small number of training images used compared to benchmark datasets, and appears to be overcome when fine tuning the network with augmented images. Moreover, these models show comparable or increased performance in the BN experiments when using augmented images indicating that the augmentations aid the softmax classifier in generalising to the data. The only experiments to show degradation in performance when fusing the image features and metadata were; 1) AlexNet when fine tuning and 2) VGG19 when fine tuning with augmented images. 

When utilising the metadata, Sensitivity, Precision, F-measure, Informedness, Markedness and MCC show a larger improvement in performance compared to using only image features, This is observed across all networks and for the majority of the configurations (BN vs FT, and augmentation). Although there is an improvement, Accuracy and Specificity show small increases relative to the other metrics in all experiments. The small change in Specificity relative to the other metrics indicate that the improvements, due to the inclusion of metadata, are due to an increase in true positives ($TP$). Specificity is the only metric without $TP$ in its definition, where all other metrics have a $TP$ in the numerator for at least one term. The small increase in Accuracy may be due to the increase in $TP$ being offset by an increase in false negative ($FN$), which is indicated by Precision, $TP / (TP+FP)$, being generally higher than Sensitivity, $TP/(TP+FN)$, where $FP$ is false positives, Fig.~\ref{fig:MacroAverageSKIN}).

\begin{figure*}[h!]
\begin{subfigure}[]{.47\textwidth}
  \centering
  \includegraphics[width=.9\linewidth]{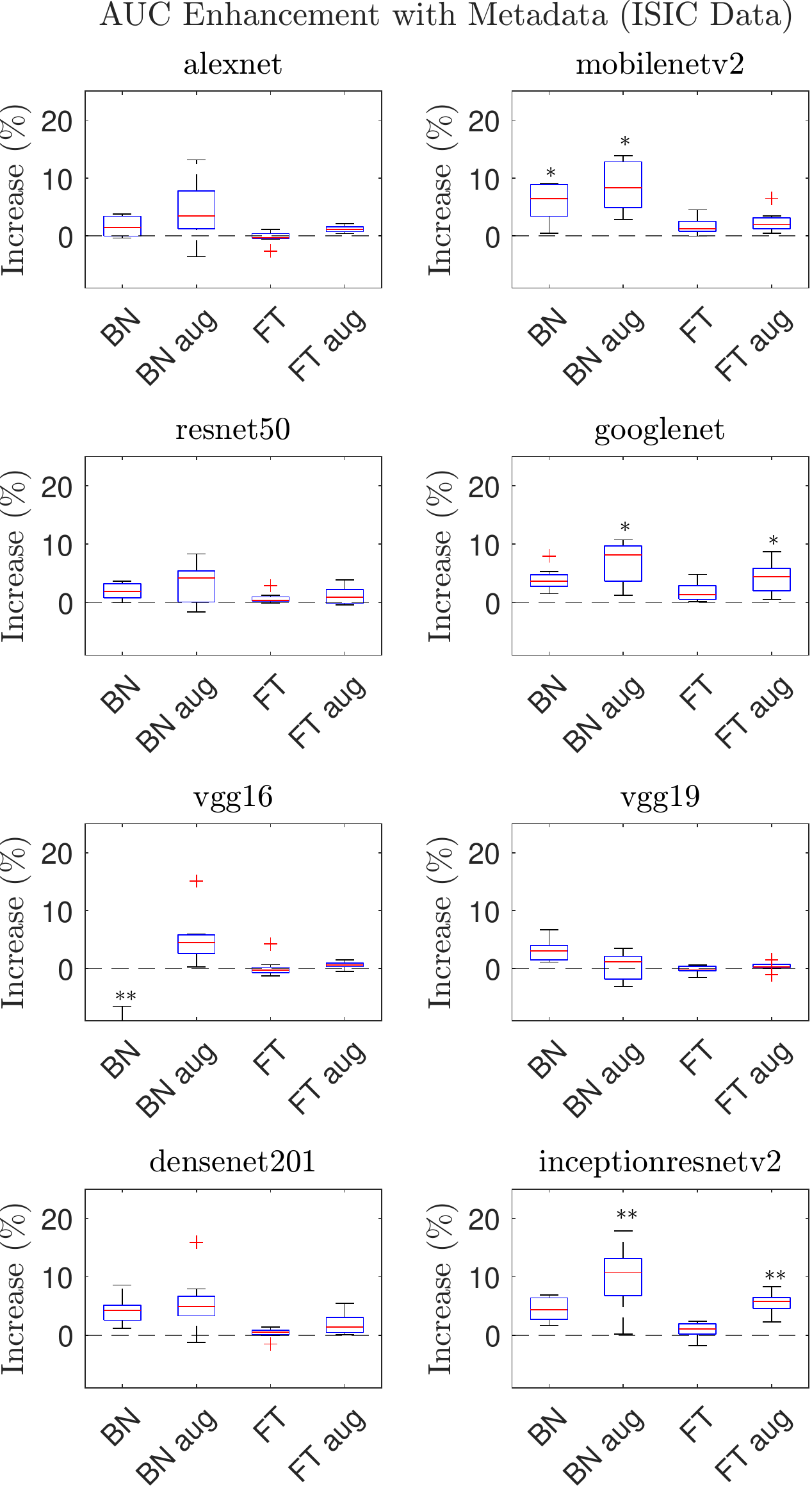}  
  \caption{ISIC}
  \label{fig:ISIC_AUC}
\end{subfigure}
\begin{subfigure}[]{.47\textwidth}
  \centering
  \includegraphics[width=.945\linewidth ]{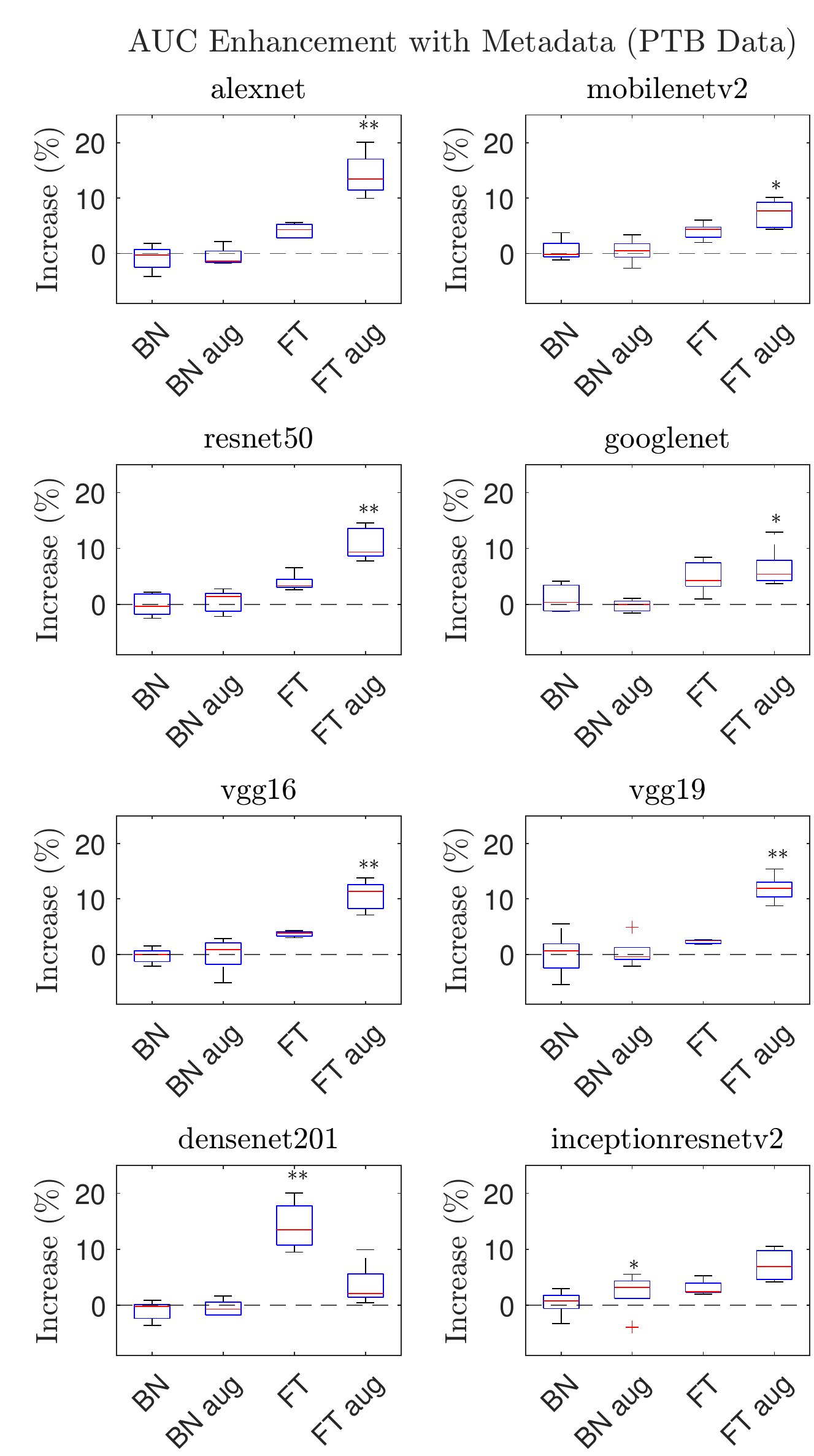}  
  \caption{PTB XL data}
  \label{fig:PTB_AUC}
\end{subfigure}
\caption{Box plots of the class-wise AUROC improvement due to the inclusion of metadata for (a) ISIC and (b) PTB XL data. Values are the difference in AUROC between models that combine image features with metadata and those based only on images features. Positive vales represent enhanced performance and negative values indicate model degradation. Results from both unprocessed and augmented images (aug) are presented. Statistically significant differences are indicated with $*$ for p$\leq$0.10 and $**$ for p$\leq$0.05. }
\label{fig:AUC}
\end{figure*}

\subsection{PTB XL Performance}
For the PTB XL data the effectiveness of transferring the ImageNet weights is different. Firstly we note that the scale of the improvement, and degradation, when using integrating image features and metadata for the PTB XL data (Fig.~\ref{fig:MacroAveragePTB}) is much larger than the performance on the ISIC data (Fig.~\ref{fig:MacroAverageSKIN}). Similar to the ISIC data, the accuracy is increased for all networks and configurations when using the metadata. 

For the BN experiments, the networks show only small changes in performance or a large degradation when using the metadata with the image features obtained from the ImageNet weights. This is observed in all metrics (other than accuracy), and although Specificity and Sensitivity are unchanged, they demonstrate a high degree of variability in improvement or degradation across the classes.  

The (FT) experiments, however, show large improvements in performance for all networks when including the metadata. For all networks except densenet201, a much larger improvement was seen when training the network with augmented images prior to combining with the metadata. In particular, the VGG architectures both showed improvements in performance metrics of up to 35\% when fine tuning the ImageNet weights prior to combining with the metadata. This highlights the potential benefits of multi modal data fusion in medical AI applications.

\begin{figure*}[t!]
\centering
\includegraphics[width=0.99\textwidth]{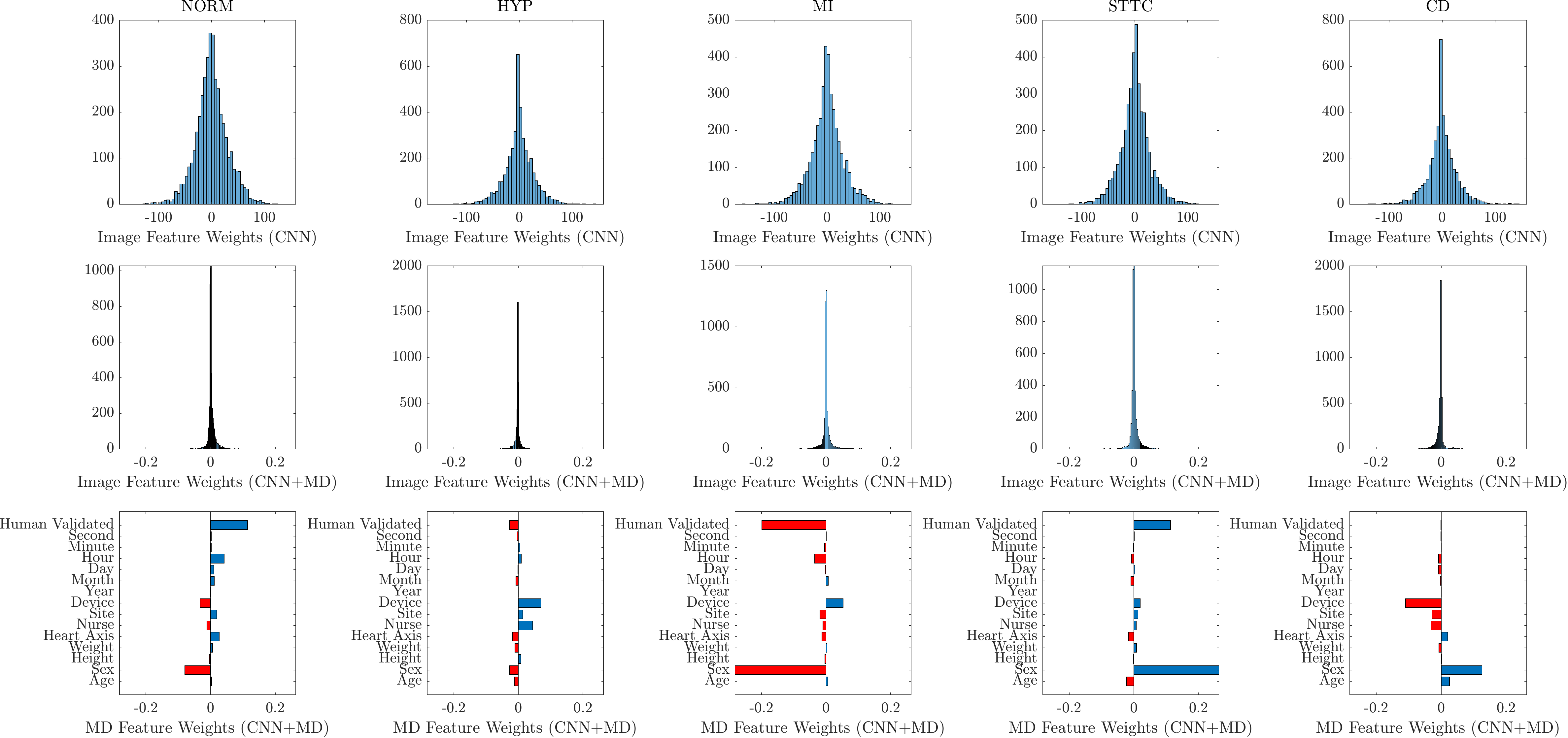}
\caption{Feature importance in the VGG16 model when fine tuned to PTB XL data with augmentation. Columns are different classes of diagnosis. Rows are (top) distribution of classifier weights for deep image features obtained from the CNN model when only using the image data, (middle) distribution of classifier weights for deep image features obtained from the CNN model when using both the image and metadata (MD), (bottom) classifier weights for the non-imaging data when combined with the deep image features.}
\label{fig:PTBexplain}
\end{figure*}

\subsection{AUROC}

We compute the receiver operator characteristic curve (AUROC) for the classifiers based only on deep image features and compare this to the AUROC of the classifiers when using both the deep image features and metadata. Similarly to the other metrics we calculate the percentage improvement in AUROC when including metadata. We represent these results as class-wise boxplots for both the ISIC and PTB XL datasets in Fig.~\ref{fig:AUC}. The boxplots represent the difference between the class-wise distribution of AUROC values, and statistically significant differences between these distortions is indicated for the p$\leq$0.10 ($*$) and p$\leq$0.05 ($**$) levels. 

In all but one case (using VGG16 as a feature extractor (BN) for the ISIC data) we observe a large increase in AUROC or minimal change when including the metadata. Both datasets also demonstrate better performance when using augmentations in the image data in general. Interestingly, the datasets differ in which transfer learning configuration yields the best results; ISIC exhibits larger improvements in AUROC when using the pretrained CNNs as feature extractors (BN) compared to fine tuning the ImageNet weights (FT); whereas for PTB XL, feature extraction had minimal impact on AUROC and large improvements on AUROC are observed when including the metadata and fine tuning the networks. This may be due to the fact that the images in the ISIC data share similar features to those in the ImageNet data set (i.e. natural images), whereas the PTB XL data are a highly specific form of data represented as images.  


\subsection{Interpretability}
We investigate how the features in the input data contribute to decisions from the classifier by looking at the input weightings $\boldsymbol{w}$ in Eq.~(\ref{eq:softmax}). Referencing to Eq.~(\ref{eq:featureFusion}), the first $d_K$ weights relate to the image features from the deep network and the last $d_{K'}$ weights relate to the metadata features. Due to the large number of models and experimental configuration in this work we provide some examples of these results and omit others for brevity. 
Figure \ref{fig:PTBexplain} demonstrates the impact of the metadata on the classifier's output with the weights for the deep imaging features (obtained from the CNNs) reduced by up to three orders of magnitude. Other networks and configurations exhibit anywhere between one and three orders of magnitude reduction in classifier weights for deep image features when the metadata is included (data not shown). The metadata features have much larger weightings than the image features when both are combined for decision making. The different classes exhibit no noticeable change in the distribution of weights for the deep image features (either with or without the metadata), though the metadata show clear discriminators for the classes. The positive weightings indicate an enhancement of feature importance for a given feature and class, whereas negative weights indicate a suppression. Figure \ref{fig:PTBexplain} shows that sex is an important feature for STTC and CD diagnosis, but irrelevant for MI. This is due to the exponent in the softmax classifier so negative weights correspond to smaller probabilities than zero weights. Interestingly, we can also see non biological factors such as measurement attributes, or potential bias in the data indicated by the importance of Device and Nurse used to collect the data. Note the relative magnitude of metadata feature weights in the classifier remain the same regardless of whether the images were augmented or not, and has no noticeable effect on the distribution of image weights for all networks (data not shown).

\begin{figure*}[t!]
\centering
\includegraphics[width=0.98\textwidth]{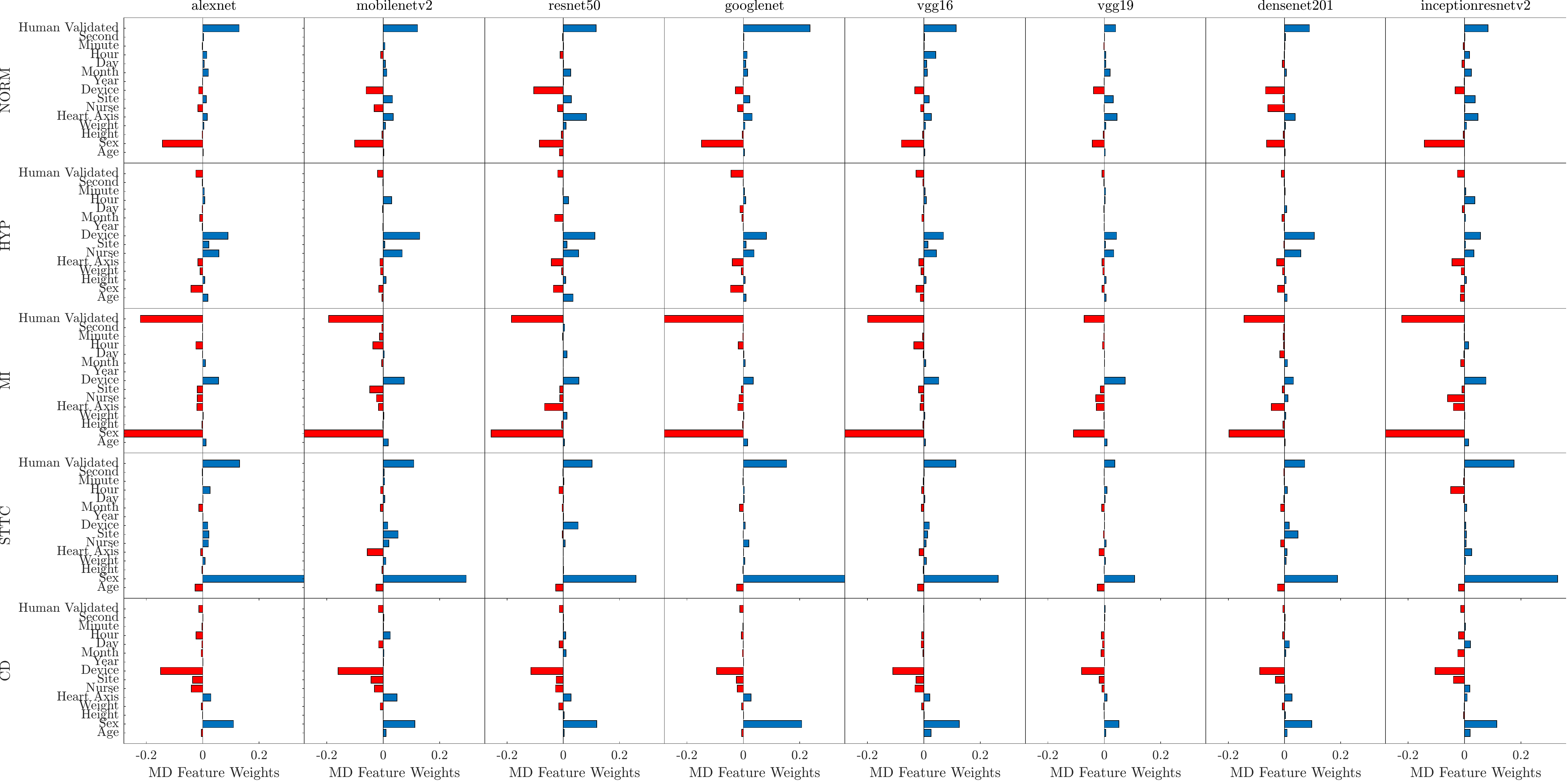}
\caption{Feature importance of metadata when combined with deep image features from a several pre-trained CNNs fine tuned with PTB XL data and image augmentation. Distributions are the classifier weights for the metadata inputs with rows depicting different classes and columns representing different CNN used for the image data. }
\label{fig:PTBexplainModels}
\end{figure*}

Figure \ref{fig:PTBexplainModels} shows the classifier weights for the metadata in the PTB XL dataset when extracting deep image features from several different CNN models separated by each class. The classifier weightings for the metadata features are similar for all the networks exhibiting the same distribution and similar magnitude within a given class. This implies the metadata features are contributing to the classification in way that is robust to the specific neural network that extracts the deep image features. This adds confidence to the importance of each metadata feature for a given diagnostic class as they are not sensitive to the combined CNN. 

\section{Conclusion}

Our results show that adding metadata to image features can significantly enhance classification performance in transfer learning. We observe that when metadata is used in the classification performance generally improves or remains the same for a range of convolutional architectures as assessed by several evaluation metrics. This indicates that this may be a general property in classification of images with deep neural networks. 

The performance enhancement depends on the type of data and the specific configuration of the transfer learning. Our experiments indicate that natural images (e.g. the ISIC data) typically exhibit greater enhancements with networks pre-trained on ImageNet directly as a feature extractor and retraining the classification layer; whereas non-natural image representation of other data (e.g. PTB XL) benefit from fine tuning the network ImageNet weights. This may arise from the similarity of characteristics in the target data with the ImageNet data, ISIC images are photos of skin lesions so may contain more comparable properties than the scalograms for the PTB XL data. This may be a general discriminator for applying transfer learning in domains with natural or non-natural images though more investigation is needed.

A barrier to the usage of metadata for data mining tasks is its heterogeneity and lack of harmonisation. Standardisation of metadata, particularly in healthcare, has become an increasingly important area due to its potential for data mining, long term curation, and compliance with the FAIR (findable, Accessible, Interoperable and Reusable) principles. Such standards will reduce the need for data cleaning and preprocessing of metadata prior to applications, making this framework readily applicable in different areas. Metadata will need to follow the same ethical and security procedures as the associated data, for example medical imaging requiring anonymisation prior to distributing. Further investigation in to these areas is required, though outside the scope of this work where we have presented a framework for utilising imaging and non-imaging data for medical applications. This methodology can be applied to other domains that use imaging data. 

It is worth highlighting that these improvements come at a negligible additional cost in computation time, and therefore are a practical method for other applications. The training time for the softmax classifier when using the combined data is comparable to when using the image features alone, typically 10s of seconds, for the entire dataset. This is insignificant compared to extraction image features from a pre-trained CNN, 10$^{3}$ - 10$^{4}$ seconds, or training the network weights, $\geq$10$^{4}$ seconds depending on the model and number of epochs. The low time cost makes this a practical extension of current methods where metadata are available, for example medical applications.

\section*{Acknowledgment}
This work was funded by the Department of Business,
Engineering and Industrial Strategy through the 
national measurement system. 
The author would like to thank Elizabeth Cooke (NPL) for her comments on the manuscript. 

\bibliographystyle{IEEEtran}
\bibliography{refs}

\end{document}